\documentclass[journal]{IEEEtran}

\usepackage{amssymb,amsthm}
\usepackage{amsfonts,amsmath,bbm}
\usepackage{latexsym}
\usepackage[mathscr]{euscript}
\usepackage{algorithm2e}
\usepackage{graphics}
\usepackage{graphicx}
\usepackage{cite}
\usepackage{cases}

\newcommand{\beq}{\begin{equation}}
\newcommand{\eeq}{\end{equation}}

\newcommand{\beqq}{\begin{equation*}}
\newcommand{\eeqq}{\end{equation*}}
\newcommand{\ei}{\end{itemize}}
\newcommand{\bi}{\begin{itemize}}

\newtheorem{definition}{Definition}

\newtheorem{prop}{Proposition}

\newtheorem{corol}{Corollary}

\theoremstyle{remark}

\newcommand{\mb}{\mathbf}

\newcommand{\ds}{\displaystyle}

\newcommand{\argmax}[1]{\arg{\hbox{$\underset{#1}{\max}\,$}}}

\begin{document}

\title{Spectrum Coordination in Energy Efficient Cognitive Radio Networks}

\author{Majed~Haddad\authorrefmark{1}
\thanks{Copyright (c) 2013 IEEE. Personal use of this material is permitted. However, permission to use this material for any other purposes must be obtained from the IEEE by sending a request to pubs-permissions@ieee.org.},~Yezekael~Hayel\authorrefmark{2} and~Oussama~Habachi\authorrefmark{2} \\
\authorrefmark{1}INRIA Sophia-Antipolis, Sophia-Antipolis, France\\
\authorrefmark{2}CERI/LIA, University of Avignon, Avignon, France
}

\maketitle

\begin{abstract}
Device coordination in open spectrum systems is a challenging problem, particularly since users experience varying spectrum availability over time and location. In this paper, we propose a game
theoretical approach that allows cognitive radio
pairs, namely the primary user (PU) and
the secondary user (SU), to update their transmission powers and frequencies
simultaneously. Specifically, we
address a Stackelberg game model in which individual users attempt to hierarchically access to the wireless spectrum while maximizing their energy efficiency.
A thorough analysis of the existence, uniqueness and characterization of the Stackelberg equilibrium is conducted. In particular, we show that a spectrum coordination naturally occurs when both actors in the system decide sequentially about their powers and their transmitting carriers. As a result, spectrum sensing in such a situation turns out to be a
simple detection of the presence/absence of a transmission on each
sub-band. We also show that when users experience very different channel gains on their two carriers, they may choose to transmit on the same carrier at the Stackelberg
equilibrium as this contributes enough energy efficiency to outweigh the interference
degradation caused by the mutual transmission. Then, we provide an algorithmic analysis on how the PU and
the SU can reach such a spectrum coordination using an appropriate
learning process. We validate our results through extensive simulations and compare the proposed algorithm to some typical scenarios including the
non-cooperative case in \cite{meshkati-jsac-2006} and the
throughput-based-utility systems. Typically, it is shown that the
proposed Stackelberg decision approach optimizes the energy efficiency
while still maximizing the throughput at the equilibrium.
\end{abstract}


\section{Introduction}

Cognitive radio technology has been proposed first to increase the throughput
of the mobiles for the next generation of wireless technologies
\cite{Hossain2009}. This enhancement is possible with an efficient use of the
wireless spectrum and specifically spectrum holes. Indeed, PUs that
have a specific and licensed access to the spectrum let part of the spectrum
unused in different time and geographic location. Many works have been done
for optimizing the behavior of SUs in cognitive radio networks (CRNs),
see \cite{Haykin05} for a survey. However, most of previous works
are focused on spectrum sharing \cite{MajedIET08} or CRN
and interference avoidance \cite{Majed-PIMRC-08}. Consequently, the energy
efficiency aspect in this setting was largely ignored. Green communications
are attracting growing attention due to various economical and environmental
reasons. This has led research community to focus more to reduce energy
consumption by introducing enhanced networking technologies
\cite{Commag11Green}, \cite{EE12GT}. Motivated by the limited battery life of mobile terminals in spite of the
transmission rate, green networking have spurred great interest and
excitement these recent years. In the literature, energy efficient power
control game has been first proposed by Goodman \emph{et al.} in
\cite{goodman-pcomm-2000} for flat fading channels and re-used by
\cite{meshkati-jsac-2006} for multi-carrier code-division multiple
access (CDMA) systems and \cite{Zappone-EE-CDMA-relay-11} for relay-assisted DS/CDMA. Most of these works do not consider the
cognitive radio technology and therefore the capabilities of the secondary
users.

In CRN, interference management is very
important since the interference due to spectrum-sharing can
significantly degrade the overall performance. In the existing
work, various resource allocation methods are proposed to
either improve energy efficiency or alleviate interference.
However, very little research has addressed their joint interaction.
In \cite{Buzzi-11-Game-EE-CDMA}, the authors considered that primary and secondary users' signals coexist in the same frequency band, and the transmit powers of the SUs are constrained so that the interference from the whole secondary network to each PU does not exceed a prescribed threshold. They formulate the problem using a non-cooperative power control game and proved the existence of a unique Nash equilibrium (NE). \cite{Bacci-13-Game-EE-OFDMA} provides an energy efficient game perspective to the problem of contention-based synchronization in orthogonal frequency-division multiple access (OFDMA) communication systems. Each user trades off its available resources so as to selfishly maximize its own revenue (in terms of probability of correct detection) while saving
as much energy as possible and satisfying quality-of-service (QoS) requirements (in terms of probability of false alarm and timing
estimation accuracy). In \cite{Rasti-11-EE-Pareto}, the authors study the gradual removal problem in energy efficient wireless networks. That is, any transmitting user whose required transmit power for reaching its target-SIR exceeds its maximum power is temporarily removed, but resumes its transmission if its required transmit power goes below a given threshold obtained in a distributed manner. Thus all transmitting users reach their target rate consuming the minimum aggregate transmit power.

We consider in this paper a hierarchical (Stackelberg) game model of a CRN in which the PU is the leader and the SU is the follower of the game. It is noteworthy that in our paper we consider the \textit{spectrum underlay} concept in which the PU experiences interference from the SU. Most of the current work has been focusing on
the spectrum sharing between cognitive radio pairs, where
cognitive radio nodes dynamically detect spectrum \emph{holes} of primary spectrum users and
opportunistically utilize them in frequency and time \cite{Haykin05}.
We formally prove that the
hierarchical structure of the game induces a spectrum coordination between the different components of the network in such a way that they transmit on distinct
carriers. This coordination property across the multiple interfering devices is particularly appealing not only from an implementation perspective, but also due to its low complexity,
smaller overhead, and ability for radio resource management (see as an example \cite{Zhao-coordination-Dyspan-2005} for open spectrum ad-hoc networks, \cite{Gesbert-MIMO-cooperation-jsac-tuto-10} for multi-cell MIMO systems and \cite{Network-coordination-Valenzuela-Wcommag-06} for cellular downlink networks).

There are many motivations for studying wireless networks with hierarchical
structures, but the most important ones are to improve the network efficiency
and modeling aspect. The Stackelberg game has been firstly proposed in
economic problem and also in biology for modeling optimal behaviors against
nature \cite{Tirole91}. It is in fact a mechanism for wireless
networks in which some wireless nodes have the priority to access the medium whereas some other nodes are equipped with cognitive sensors like in CRN (see \cite{cn/Kim12} which is one of the first reference which addressed a multi-leader and multi-follower game theoretic model for CR spectrum sharing networks). This is also a natural setting for heterogeneous wireless networks due to the absence
of coordination among the small cells and between small cells and macro cells \cite{Femto08Survey,Majed_wiopt14}. At the core lies the idea that the utility of the leader obtained at the Stackelberg equilibrium is higher than his utility
obtained at the NE when the two users play simultaneously. This is
due to the Stackelberg mechanism in which the leader anticipates the
follower's action. It
has been proved in \cite{samson-twc09} that this result is also true for the
follower. The goal is then to find a Stackelberg equilibrium in this
two-step game \cite{bilevel05}.

The original contributions of this paper are threefold: \bi
\item Introducing hierarchy concept in power control game for energy
 efficient multi-carrier systems,
\item Characterizing completely and analytically the Stackelberg equilibrium
 and compare the results obtained in the proposed hierarchical game with
 those obtained in the non-cooperative game in \cite{meshkati-jsac-2006},
\item Our main result is that we \textbf{\emph{always}} obtain an equilibrium
 (contrary to the work addressed in \cite{meshkati-jsac-2006}) where, for
 the most general cases, the two users transmit on \textbf{\emph{distinct}}
 carriers delivering a binary channel assignment.
\ei

The organization of the paper is the following. First, we introduce in Section
\ref{sec:model} the CRN context and the different decision makers of the
system. In Section \ref{sec:net-energy-eff}, we define the energy efficiency framework which is used throughout the paper and present the game theoretic model in Section \ref{sec:game}. Next, in Section \ref{sec:SE}, we characterize the Stackelberg equilibrium by providing a thorough analysis on the existence and uniqueness of such an equilibrium. Having these results, we then
address the important property of spectrum coordination in
Section \ref{sec:coordin}. Section \ref{sec:implem} provides an analysis of the implementation issues including a learning algorithm that ensures convergence to the Stackelberg equilibrium in \ref{sec:learn} and the sensing issue in \ref{sec:sens}. Section \ref{sec:sim} illustrates
some numerical results and Section \ref{sec:conc} concludes the paper.

\section{The Cognitive Radio System Model}\label{sec:model}

We consider a network composed of a PU (or leader
-- indexed by $1$), having the priority to access the medium, and a SU (or follower -- indexed by $2$) that accesses the medium after observing the action of the PU subject to mutual interference. We assume slotted transmissions (over carriers) for both the PU
and the SU. The equivalent baseband signal received by the base
station can be written as
\begin{equation}
y_k = \displaystyle h_{1k} x_{1k} + h_{2k} x_{2k} + z_k, \,\,\, \mathrm{for}\, \, k = 1,2
\label{eq:received-signal}\end{equation} where $h_{nk}$ stands for the block fading process of
user $n$ on the sub-band $k$, $x_{nk}$ is the signal transmitted by user $n$
on the sub-band $k$ and $z_k$ is the additive Gaussian noise at the $k$th
sub-band. We denote by $g_{nk} = |h_{nk}|^2$ the fading channel gain which is
assumed to stay constant over each block fading length (i.e., coherent
communication). We statistically model the channel gains $\textbf{g}_{n}$ to
be i.i.d. distributed over the Rayleigh fading coefficients.
The signal transmitted
$x_{nk}$ can be further written as $x_{nk} = \sqrt{p_{nk}}s_{nk}$ where $p_{nk}$
and $s_{nk}$ are the transmit power and data of user $n$. We thus have
$\mathbb{E}\left\{|x_{nk}|^2\right\} = p_{nk}$. The additive Gaussian noise
$z_k$ at the receiver is i.i.d. circularly symmetric and
$z_k\sim\mathcal{CN}(0,\sigma^2)$ for $k= 1,2$. For any user $n \in \{1,2\}$
the received signal-to-noise plus interference ratio (SINR) over carrier $k$
is expressed as \beq\label{eq:gamma-mc}
\displaystyle \gamma_{nk}=\frac{g_{nk} p_{nk}}{\sigma^2+ \displaystyle\sum_{\substack{m=1 \\
m\neq n}}^2 g_{mk} p_{mk}}:=p_{nk} \widehat{h}_{nk}. \eeq
In the remainder, we will define the ratio between the SINR $\gamma_{nk}$ and the transmission power $p_{nk}$ by the \emph{effective channel gain} $\widehat{h}_{nk}$.
It follows from the
above SINR expression that the strategy chosen by a user (i.e., the power vector $\mathbf{p_{n}}=(p_{n1},p_{n2})$) may affect the
performance of the other user in the network through multiple-access
interference reflected by the effective channel gain.

\section{Network Energy Efficiency Analysis}
\label{sec:net-energy-eff}

Our system model is based on the seminal paper \cite{goodman-pcomm-2000}
that defines the energy efficiency framework. In order to formulate the power control
problem as a game, we first need to define a utility function suitable for
data applications.
Increasing the transmit power clearly favors the packet success rate and
therefore the throughput. However, as the packet success rate tends to one,
further increasing the power can lead to marginal gains in terms of
throughput regarding the amount of extra power used. The following utility
function allows one to measure the corresponding trade-off between the
transmission benefit (total throughput over both carriers) and cost (total
power over both carriers)\footnote{Notice that although this is not a restriction
of the proposed analysis and for the sake of simplicity in the notations, we do not consider the circuit power needed to operate user $n$ in the definition of the consumed power in the denominator of Eq.~(\ref{eq:util-mc}).}:
\beq\label{eq:util-mc}
u_n(\mb{p_1},\mb{p_2})=\frac{\displaystyle R_n \cdot (
f(\gamma_{n1})+f(\gamma_{n2}))}{p_{n1}+p_{n2}}
\eeq
where $R_n$ is the transmission rate of user $n$ and $f(\cdot)$ is an
increasing, continuous and S-shaped \emph{efficiency function} which measures the packet success rate. A more detailed discussion of the efficiency function can be found in \cite{meshkati-spmag-2007}. The utility function $u_n$ that has units of bits per joule perfectly captures the trade-off between throughput and battery life and is particularly suitable for applications where energy efficiency is crucial such as sensors and mobiles terminals.

\section{The Game Theoretic Framework}
\label{sec:game}

\subsection{The non-cooperative game problem}

An important solution concept of the game under consideration is the NE \cite{nash50}, which is a fundamental concept in
non-cooperative strategic games. It is a vector of strategies (or actions in
our case) $\mb{p}^{NE} = \{\mb{p_1}^{NE},\mb{p_2}^{NE}\}$, one for each player, such
that no player has incentive to unilaterally deviate, i.e.,
$u_n(\mb{p_n}^{NE},\mb{p}_{-n}^{NE})\geq u_n(\mb{p_n},\mb{p}_{-n}^{NE})$ for all action $\mb{p_n}
\neq \mb{p_n}^{NE}$, where the $-n$ subscript on vector $\mb{p}$ stands for
"except user $n$". In \cite{goodman-pcomm-2000}, authors showed that, under certain conditions, the
NE of the game with utility (\ref{eq:util-mc}) exists.

\subsection{The hierarchical game formulation}\label{sec:Stack}

In this work, we consider a Stackelberg game framework in which the PU decides first his power control vector $\mathbf{p_{1}}$
and based on this, the SU will adapt his power control
vector $\mathbf{p_{2}}$.

\begin{definition} ({\bf Stackelberg equilibrium}):
\emph{A vector of actions
$\mathbf{\widetilde{p}}=(\mathbf{\widetilde{p}_1},\mathbf{\widetilde{p}_2})=(\widetilde{p}_{11},\widetilde{p}_{12},\widetilde{p}_{21},\widetilde{p}_{22})$
is called Stackelberg equilibrium (SE) if and only if:
$$
\mathbf{\widetilde{p}_{1}} =\argmax{\mathbf{p_1}} u_1(\mathbf{p_1},\mb{\overline{p}_2}(\mathbf{p_1})),
$$
where
$$
\forall \mathbf{p_1},\quad \mb{\overline{p}_2}(\mathbf{p_1})=\argmax{\mathbf{p_2}}u_2(\mathbf{p_1},\mathbf{p_2}),
$$
and $\mathbf{\widetilde{p}_2}=\mb{\overline{p}_2}(\mathbf{\widetilde{p}_1})$.}
\end{definition}

A SE can be determined using a bi-level approach \cite{bilevel05}. Given
the action of the PU, we compute the best-response function of the
SU (the function $\overline{p}_2(\cdot)$), i.e., the action of
the SU which maximizes his utility given the action of the
PU. This best-response function is characterized by using a result
from \cite{meshkati-jsac-2006} which depends on the PU's power control on carrier $k$ through the following expression:
$$
\forall k \in \{1,2\},\quad
\widehat{h}_{2k}(p_{1k})=\frac{\gamma_{2k}}{p_{2k}}=\frac{g_{2k}}{\sigma^2+g_{1k}p_{1k}}.
$$

\section{Characterization of the Stackelberg Equilibrium}\label{sec:SE}

In order to determine the SE, a standard approach is to consider a backward induction technique. Then, we first determine the best-response function of the SU depending on the action of the PU. This result comes directly from
Proposition 1 of \cite{meshkati-jsac-2006}. For making this paper
sufficiently self-contained, we review here the latter proposition.

\subsection{The secondary user's power control vector}
\begin{prop}\label{prop:secondary user-power}\textbf{(Given in \cite{meshkati-jsac-2006})}

Given the power control vector $\mathbf{p_1}$ of the PU, the
best-response function of the SU is given by
\begin{equation}\label{eq:secondary user}
\overline{p}_{2k}(\mathbf{p_1})= \left\{\begin{array}{lr}\displaystyle
\frac{\gamma^{*}(\sigma^2+g_{1k} p_{1k})}{g_{2k}},&\, \mbox{for} \,\, k = L_2(\mathbf{p_1}),\\
0,& \, \mbox{for all}\,\, k \neq L_2(\mathbf{p_1})
\end{array}
\right.
\end{equation}
with $L_2(\mathbf{p_1})=\argmax{k} \widehat{h}_{2k}(p_{1k})$ and $\gamma^*$ is
the unique (positive) solution of the first order equation
\begin{equation}\label{eq:gamma*}
x\,f^{\prime}(x)=f(x).
\end{equation}
\end{prop}

Equation (\ref{eq:gamma*}) has a unique solution if the efficiency function
$f(\cdot)$ is sigmoidal \cite{rodriguez-globecom-2003}, and we will use this
assumption throughout our paper.

Proposition \ref{prop:secondary user-power} claims that there are two regions
depending on the PU's power control which yields different best-response functions for
the SU. Below, we define the two regions:
\begin{eqnarray}\label{eq:A}
\mathcal{A}&=&\ds\left\{(p_{11},p_{12})|\widehat{h}_{22} \geq \widehat{h}_{21}\right\}\nonumber\\
&=&\ds\left\{(p_{11},p_{12})|p_{12} \leq p_{11}\frac{g_{11} g_{22}}{g_{12} g_{21}} + \sigma^2
\frac{(g_{22}-g_{21})}{g_{12} g_{21}}\right\}
\end{eqnarray}
and
\begin{eqnarray}\label{eq:B}
{\mathcal{B}}&=&\ds\left\{(p_{11},p_{12})|\widehat{h}_{22} < \widehat{h}_{21}\right\}\nonumber\\
&=&\ds\left\{(p_{11},p_{12})|p_{12} > p_{11}\frac{g_{11} g_{22}}{g_{12} g_{21}} + \sigma^2
\frac{(g_{22}-g_{21})}{g_{12} g_{21}}\right\}.
\end{eqnarray}

\subsection{The primary user's power control vector}

So far, we have seen that the best-response function of the SU is to use only one carrier, the one with the best effective channel gain.
Let us now study the optimal power control for the PU knowing the
best-response function of the SU. The following
proposition, which is our first main result, gives the existence and uniqueness of the optimal power control of the PU at the SE knowing the
best-response function of the SU. Notice that uniqueness of the SE is a desirable property for a
Stackelberg game. If there exists exactly one equilibrium, we can predict the
equilibrium strategy of the players and resulting performance of the system.

\begin{prop} \textbf{(First main result) Existence and uniqueness of the PU's power control at the SE}\label{prop:existence-se}\\
There exists a unique power control vector $\widetilde{\mb{p}}_{1}=( \widetilde{p}_{11} \widetilde{p}_{12})$ for the PU which maximizes his energy efficiency over Region $\mathcal{A}$. It is defined by:
\begin{equation*}\label{eq:primary userA}
\widetilde{p}_{12}=0,\,\, \text{and} \,\, \widetilde{p}_{11}= \left\{\begin{array}{lr}\displaystyle
\frac{\sigma^2\gamma^*}{g_{11}},&\, \mbox{if} \,\ \ds\frac{g_{22}}{g_{21}}\geq\frac{1}{1+\gamma^*},\\
\ds\frac{\sigma^2(g_{21}-g_{22})}{g_{11}g_{22}},& \, \mbox{otherwise}.
\end{array}
\right.
\end{equation*}

There exists a unique power control vector $\widetilde{\mb{p}}_{1}=( \widetilde{p}_{11}, \widetilde{p}_{12})$ for the PU which maximizes his energy efficiency over Region ${\mathcal{B}}$. It is defined by:
\begin{equation*}\label{eq:primary userB}
\widetilde{p}_{11}=0\,\, \mbox{and} \,\, \widetilde{p}_{12}= \left\{\begin{array}{lr}\displaystyle
\frac{\sigma^2\gamma^*}{g_{12}},&\, \mbox{if} \,\, \ds\frac{g_{22}}{g_{21}}\leq1+\gamma^*,\\
\ds\frac{\sigma^2(g_{22}-g_{21})}{g_{12}g_{21}},& \, \mbox{otherwise}.
\end{array}
\right.
\end{equation*}
\end{prop}

For the clarity of the exposition, this proposition is proven in Appendix \ref{app:exis}.

This result combined with the result of Prop. \ref{prop:secondary user-power} yields the existence of a SE.

\begin{corol}\label{corol:stack}
At the Stackelberg equilibrium, when the channel gains of the SU satisfy \begin{eqnarray}\label{chfol}
\frac{1}{1+\gamma^*}\leq \frac{g_{21}}{g_{22}}\leq1+\gamma^*,
\end{eqnarray}
the power control vector $\mathbf{\widetilde{p}_1}$ which maximizes the PU's utility is \textbf{unique} and is given by
\begin{equation}\label{eq:primary user}
\widetilde{p}_{1k}= \left\{\begin{array}{lr}\displaystyle
\frac{\sigma^2\gamma^*}{g_{1k}},& \mbox{for} \quad k=\widetilde{k},\\
0,& \mbox{for all}\quad k \neq \widetilde{k},
\end{array}
\right.
\end{equation}
where $\widetilde{k}$ denotes the "best"
carrier of the PU, i.e., $\widetilde{k} = \argmax{k} g_{1k}$.
\end{corol}

\begin{proof}
The proof makes use of results from Prop. \ref{prop:existence-se} for the PU's power control in Region $\mathcal{A}$ and ${\mathcal{B}}$.
We have
that the utility of the PU within Region $\mathcal{A}$ is maximized when
$p_{12}=0$, yielding
\beq\label{eq:p11A}
\max_{p_{11},p_{12}}u_1^{\mathcal{A}}(p_{11},p_{12})=\max_{p_{11}}u_1^{\mathcal{A}}(p_{11},0)=\max_{p_{11}}\frac{R_1f(\frac{g_{11}p_{11}}{\sigma^2})}{p_{11}}
\eeq
which implies that the maximum utility over Region $\mathcal{A}$ is given by $\widetilde{u}_1^{\mathcal{A}}=u_1^{\mathcal{A}}(\widetilde{p}_{11},0)$.

Within Region ${\mathcal{B}}$, we have that the utility of the PU is maximized when $p_{11}=0$, yielding
\beq\label{eq:p12B}
\max_{p_{11},p_{12}}u_1^{\mathcal{B}}(p_{11},p_{12})=\max_{p_{12}}u_1^{\mathcal{B}}(0,p_{12})=\max_{p_{12}}\frac{R_1f(\frac{g_{12}p_{12}}{\sigma^2})}{p_{12}}.
\eeq
which implies that the maximum utility within Region ${\mathcal{B}}$ is $\widetilde{u}_1^{\mathcal{B}}=u_1^{\mathcal{B}}(0,\widetilde{p}_{12})$.
Combining the above results for Region ${\mathcal{A}}$ (in Eq. (\ref{eq:p11A})) and Region ${\mathcal{B}}$ (in Eq. (\ref{eq:p12B})), the maximization problem of the PU's utility becomes
\begin{equation*}
\max_{p_{11},p_{12}}u_1(p_{11},p_{12})=\max(\widetilde{u}_1^{\mathcal{A}},\widetilde{u}_1^{\mathcal{B}})=
\left\{\begin{array}{lr}\displaystyle
\widetilde{u}_1^{\mathcal{A}},&\,\, \mbox{if} \quad g_{11}\geq g_{12},\\
\widetilde{u}_1^{\mathcal{B}},& \,\, \mbox{if}\quad g_{11}< g_{12}
\end{array}
\right.
\end{equation*}
where we use the fact that $f$ is a strictly increasing function. This completes the proof.
\end{proof}

Condition \ref{chfol} means that a given
user experiences approximately the same channel characteristics over his two carriers. Note that this is typically the case when the two carriers are close enough \cite{proakis01}.
Corollary \ref{corol:stack} says that the utility of PU is
maximized when he transmits only over his \emph{best} carrier. Accordingly,
we observe that the carrier which doest not provide enough energy efficiency
to outweigh the interference degradation caused by the SU's
transmission is switched "off". Notice that this result is in contradiction with throughput-based-utility systems which lead to a water-filling power control \cite{tse-book} where only a certain number of carriers are exploited depending on the channel gains.

To resume, Prop. \ref{prop:secondary user-power} and Prop. \ref{prop:existence-se} suggest
that, at the SE, both the SU and the PU transmit on only one
carrier depending on their channel gains. In the next section, we will
show that hierarchy "pushes" users towards
coordinating their actions in such a way that they transmit on distinct
carriers.

\section{Spectrum Coordination}\label{sec:coordin}

\subsection{General result}

A necessary and sufficient condition on the SU's channel gains such that the best-response function of the SU is to transmit over a
distinct carrier than the PU is given in the following
proposition.

\begin{prop}\label{prop:stack-0}
At the Stackelberg equilibrium, if the PU transmits over only
\textbf{one carrier}, the SU transmits over \textbf{a distinct
carrier} if and only if Condition (\ref{chfol}) is satisfied.
\end{prop}

The proof of Prop. \ref{prop:stack-0} is given in Appendix \ref{app:stack-0}.
Prop. \ref{prop:stack-0} claims that Condition (\ref{chfol}) is a necessary and sufficient condition to obtain spectrum coordination. We will see in the next proposition that a spectrum coordination can occurs even if Condition (\ref{chfol}) is not satisfied. In this case, the SE is not unique as the SU obtains the same utility by choosing to transmit either on a different carrier from the PU (coordination case) or on the same carrier than the PU (non coordination case).

\begin{prop}\label{prop:coordination}\textbf{(Second main result) Spectrum Coordination}\\
Introducing hierarchy between users in a
two-carrier energy efficient power control game induces a natural coordination pattern where users have incentive to choose
their transmitting carriers in such a way that they transmit on
orthogonal channels.
\end{prop}

\begin{proof}
To show this important result, we will determine the Stackelberg equilibria
of the users depending on their channel gains. As far as the proposed hierarchical model is concerned, the
SE can be computed by considering the following possibilities:

\begin{itemize}
\item (a) If $\ds\frac{1}{1+\gamma^*}\leq\frac{g_{21}}{g_{22}}\leq1+\gamma^*$ (i.e., the SU experiences approximately the same radio conditions over his two carriers),

 \bi
 \item (i) if $\ds\frac{g_{11}}{g_{12}}<1$, then $\overline{p}_2(0,\frac{\sigma^2\gamma^*}{g_{12}})=(\frac{\sigma^2\gamma^*}{g_{21}},0)$ because $\widehat{h}_{22}=\frac{g_{22}}{\sigma^2(1+\gamma^*)}$ and
$\widehat{h}_{21}=\frac{g_{21}}{\sigma^2}$. \\
The SE is then given by:
\beq
 \left(\widetilde{p}_{11},\widetilde{p}_{12},\widetilde{p}_{21},\widetilde{p}_{22}\right)=\left(0,\frac{\gamma^*\sigma^2}{g_{12}},\frac{\gamma^*\sigma^2}{g_{21}},0\right),
 \eeq
\item (ii) otherwise, $\ds\frac{g_{11}}{g_{12}}\geq1$, then $\overline{p}_2(\frac{\sigma^2\gamma^*}{g_{11}},0)=(0,\frac{\sigma^2\gamma^*}{g_{22}})$ because $\widehat{h}_{21}=\frac{g_{21}}{\sigma^2(1+\gamma^*)}$ and
$\widehat{h}_{22}=\frac{g_{22}}{\sigma^2}$. \\
The SE is then given by:
is
 \beq
 \left(\widetilde{p}_{11},\widetilde{p}_{12},\widetilde{p}_{21},\widetilde{p}_{22}\right)=\left(\frac{\gamma^*\sigma^2}{g_{11}},0,0,\frac{\gamma^*\sigma^2}{g_{22}}\right).
 \eeq
\end{itemize}

\item (b) If $\ds\frac{g_{21}}{g_{22}}>1+\gamma^*$ (i.e., the SU
 experiences deep fade on his second carrier compared to his first
 carrier),
 \begin{itemize}
 \item (i) if $\ds\frac{g_{11}}{g_{12}}<1$, then the Stackelberg
 equilibrium is
 \beq
 \left(\widetilde{p}_{11},\widetilde{p}_{12},\widetilde{p}_{21},\widetilde{p}_{22}\right)=\left(0,\frac{\gamma^*\sigma^2}{g_{12}},\frac{\gamma^*\sigma^2}{g_{21}},0\right).
 \eeq
 \item (ii) otherwise, $\ds\frac{g_{11}}{g_{12}}\geq1$, the power control vector of the PU at the SE is
 \begin{equation*}
\left(\widetilde{p}_{11},\widetilde{p}_{12}\right)= \left\{\begin{array}{lr}\displaystyle
\left(\frac{\sigma^2(g_{21}-g_{22})}{g_{11}g_{22}},0\right)&\quad \mbox{if (\ref{cond1})},\\
\left(0,\frac{\sigma^2\gamma^*}{g_{12}}\right)& \quad \mbox{otherwise}.\\
\end{array}
\right.
\end{equation*}
where Condition (\ref{cond1}) is
\begin{eqnarray}\label{cond1}
\frac{g_{11}}{g_{12}}\geq\frac{f(\gamma^*)}{\gamma^*}\frac{\frac{g_{21}}{g_{22}}-1}{f(\frac{g_{21}}{g_{22}}-1)}.
\end{eqnarray}

The SU transmits on the carrier which is left idle by the PU if Condition (\ref{cond1}) is not satisfied. In this case
 $$
 \left(\widetilde{p}_{21},\widetilde{p}_{22}\right)=\left(\frac{\sigma^2\gamma^*}{g_{21}},0\right).
 $$
 If Condition (\ref{cond1}) is satisfied, we have the following
 best-response function for the SU:
 $$
 \overline{p}_2(\frac{\sigma^2(g_{21}-g_{22})}{g_{11}g_{22}},0)=\left\{(\frac{\sigma^2\gamma^*}{g_{22}},0)\,\,
 \mbox{or}\,\,(0,\frac{\sigma^2\gamma^*}{g_{22}})\right\},
 $$
 because the effective channel gains are equal for both carriers,
 i.e., $\widehat{h}_{21}=\widehat{h}_{22}=\frac{g_{22}}{\sigma^2}$.
 Then the best-response function is not unique in this case and the two
 players can use the same carrier, the first one here. As the
 SU plays after observing the action of the primary
 user, the SU can decide, for optimizing spectrum
 utilization, to transmit over the carrier left idle by the
 PU. Moreover, the SU's power is inversely
 proportional to the channel gain over the second carrier. Then,
 it is more convenient for him to transmit over this second
 carrier.
 \end{itemize}
 \item (c) If $\ds\frac{g_{21}}{g_{22}}<\frac{1}{1+\gamma^*}$ (i.e., the
 SU experiences deep fade on his first carrier compared
 to his second carrier), we have the similar results:
 \begin{itemize}
 \item (i) if $\ds\frac{g_{11}}{g_{12}}\geq1$, then the SE
 is
 \beq
 \left(\widetilde{p}_{11},\widetilde{p}_{12},\widetilde{p}_{21},\widetilde{p}_{22}\right)=\left(\frac{\gamma^*\sigma^2}{g_{11}},0,0,\frac{\gamma^*\sigma^2}{g_{22}}\right).
 \eeq
 \item (ii) otherwise, $\ds\frac{g_{11}}{g_{12}}<1$, the power control
 vector of the PU at the SE is
 \begin{equation*}
\left(\widetilde{p}_{11},\widetilde{p}_{12}\right)= \left\{\begin{array}{lr}\displaystyle
\left(0,\frac{\sigma^2(g_{22}-g_{21})}{g_{12}g_{21}}\right)&\quad \mbox{if (\ref{cond2})},\\
\left(\frac{\sigma^2\gamma^*}{g_{11}},0\right)& \quad \mbox{otherwise}.\\
\end{array}
\right.
\end{equation*}
where Condition (\ref{cond2}) is
\begin{eqnarray}\label{cond2}
\frac{g_{11}}{g_{12}}<\frac{\gamma^*}{f(\gamma^*)}\frac{f(\frac{(g_{22}-g_{21})}{g_{21}})}{\frac{(g_{22}-g_{21})}{g_{21}}}.
\end{eqnarray}

 The SU transmits on the carrier which is left idle by
 the PU if Condition (\ref{cond2}) is not satisfied. In
 this case
 $$
 \left(\widetilde{p}_{21},\widetilde{p}_{22}\right)=\left(0,\frac{\sigma^2\gamma^*}{g_{22}}\right).
 $$
 If Condition (\ref{cond2}) is satisfied, we have the following
 best-response function for the SU:
 $$
 \overline{p}_2(0,\frac{\sigma^2(g_{22}-g_{21})}{g_{12}g_{21}})=\left\{(0,\frac{\sigma^2\gamma^*}{g_{21}})\,\,
 \mbox{or}\,\,(\frac{\sigma^2\gamma^*}{g_{21}},0)\right\},
 $$
 because the effective channel gains are equal for both carriers,
 i.e., $\widehat{h}_{21}=\widehat{h}_{22}=\frac{g_{21}}{\sigma^2}$.
 Then the best-response function is not unique in this case and the two
 players can use the same carrier, the second one here. In this
 particular case, the SU can decide to transmit over
 the first carrier in order to optimize the spectrum utilization.
 Again, as the SU's power is inversely proportional to
 the channel gain on the first carrier, it is more convenient for
 him to transmit over this first carrier.
 \end{itemize}
\end{itemize}
\end{proof}
\vspace{-0.5cm}
Having treated the case of spectrum coordination, let us now present a particular case (on the fading channel ) where the two players gain by transmitting on the same carrier at the SE.

\subsection{Extreme Case}

In a Stackelberg game, if the leader decides to play a Nash action,
then the follower plays the Nash action too as it is the best-response function
to the Nash action. Then, depending on the ratio $\ds\frac{g_{11}}{g_{12}}$, it could be interesting for the PU to
transmit over the same channel than the SU. We will show in the next proposition that this case can appear, essentially when the target SINR at the SE is very low, i.e., $\gamma^*<1$ and with some conditions on the channel gains.

\begin{prop}\label{stackNash}
At the Stackelberg equilibrium, in Region $A$ (resp. Region
${\mathcal{B}}$), if $\gamma^*<1$, both the PU
and the SU transmit on the first (resp. second) carrier if
\beq\label{eq:ext-case}
\frac{g_{n1}}{g_{n2}}\geq\frac{1}{1-\gamma^*},\quad (\mbox{resp.} \quad \frac{g_{n1}}{g_{n2}}\leq1-\gamma^*),\quad \mbox{for} \,\,n \in \{1,2\}.
\eeq
\end{prop}

The proof of Prop. \ref{stackNash} is given in Appendix \ref{app:stackNash}. Prop. \ref{stackNash} claims that the probability of extreme case turns out to be the probability of no coordination between users. Specifically, in the extreme case of Region ${\mathcal{A}}$ ($\frac{g_{11}}{g_{12}}\geq1/(1-\gamma^*)$), the
PU decides to transmit on the same carrier (second one here) as the
second carrier is much better that the first one. In the extreme
case of Region ${\mathcal{B}}$ ($\frac{g_{11}}{g_{12}}\leq1-\gamma^*$), the channel gain is
very \emph{bad} on the second carrier with respect to the one on the first
carrier and then both users choose to transmit on the first carrier. Note
that, in this case, the SU and the PU transmit over the
same carrier using an optimal power control given by the Stackelberg model
proposed in \cite{samson-twc09}.

{\normalsize
Notice that, in the case of Rayleigh fading channels, the probability
of being in the extreme case is given by:}

{\scriptsize
\begin{eqnarray*}\label{eq:probab}
\psi(\gamma^*)&=& Pr\left\{\frac{g_{11}}{g_{12}}\geq\frac{1}{1-\gamma^*}\right\}\cdot Pr\left\{\frac{g_{21}}{g_{22}}\geq\frac{1}{1-\gamma^*}\right\}
\\&&
+ Pr\left\{\frac{g_{11}}{g_{12}}\leq 1-\gamma^*\right\}\cdot Pr\left\{\frac{g_{21}}{g_{22}}\leq 1-\gamma^*\right\}\\\\
&=& \displaystyle \left[\int_0^{\infty} \int_{\frac{y}{(1-\gamma^*)}}^{\infty} e^{-(x+y)} dx dy\right]^2
\\&&
+ \left[\int_0^{\infty} \int_0^{(1-\gamma^*)y} e^{-(x+y)} dx dy\right]^2= \displaystyle 2\cdot \left(\frac{\gamma^*-1}{\gamma^*-2}\right)^2.
\end{eqnarray*}
}

\begin{figure}[t]
\vspace*{-0cm}
\hspace*{-0cm}
\centering
\includegraphics[height =4.5cm,width=8.5cm]{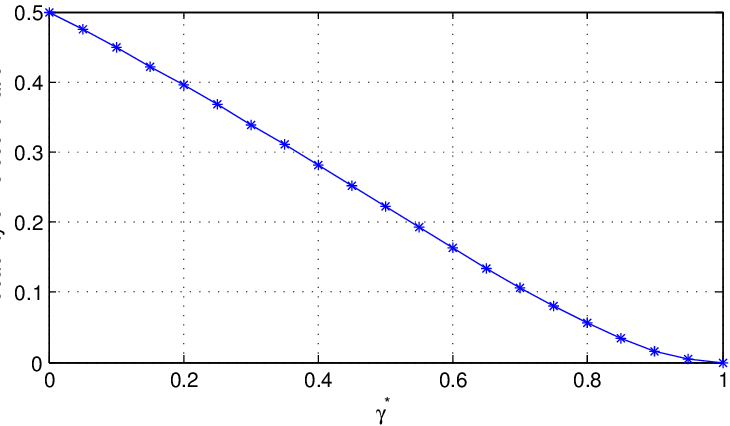}
\vspace*{-0cm}
\caption{The probability of being in the extreme case (or the probability of no coordination) considering Rayleigh fading channels.}
\label{fig:probab}
\end{figure}

Figure \ref{fig:probab} depicts the probability of being in the extreme
case -- which is the probability of no coordination -- when $\gamma^*<1$. It is shown that the probability of being in
the extreme case is always lower than $0.5$. As $\gamma^*$
increases, the extreme region shrinks resulting in a decrease of the probability of no coordination.

\begin{figure}[t]
\vspace*{-0cm}
\hspace*{-2.7cm}
\centering
\includegraphics[height = 18cm,width=11.7cm]{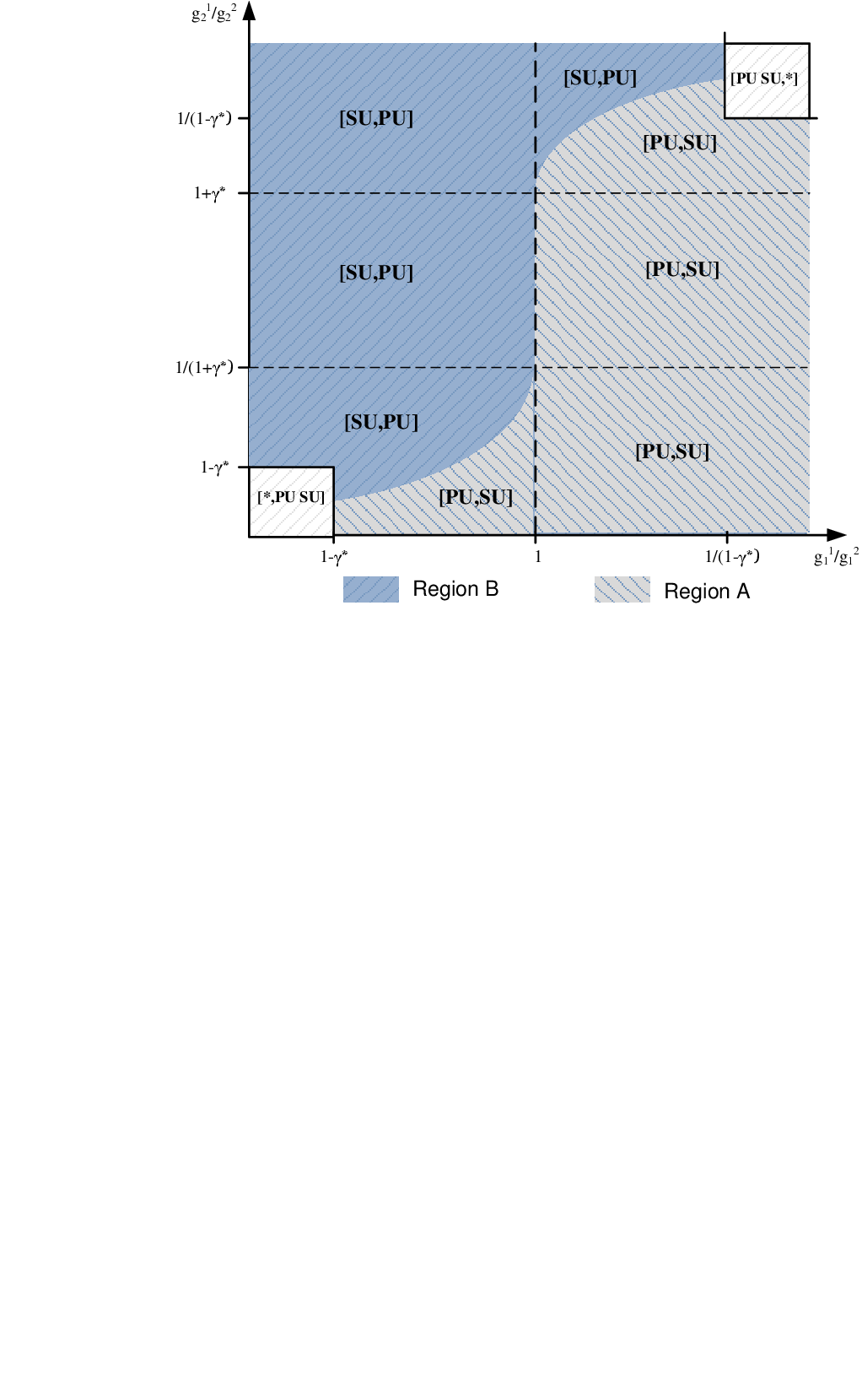}
\vspace*{-10.5cm}
\caption{Stackelberg equilibrium regions for the case of two users and two carriers. The point $[SU,PU]$ means that the PU transmits over the second carrier and the SU over the first one.}
\label{fig:SE}
\end{figure}

A global overview of the occupation of the carriers at the SE, as function of the ratios $\ds\frac{g_{21}}{g_{22}}$ and
$\ds\frac{g_{11}}{g_{12}}$ is depicted in Figure \ref{fig:SE}. It is shown the main
contributions of the paper, namely
\bi
\item we have proved the \emph{existence} and \emph{uniqueness} of an
equilibrium when a user can observe the action of the other user before deciding
his own action, whatever the channel gains are. This result is not true in
the case when the two users play a NE (see for instance
\cite{meshkati-jsac-2006}),
\item although we have formulated the problem of
energy efficiency maximization by allowing that a carrier could be
shared by both users, we have obtained a \emph{spectrum coordination} pattern in which, to refrain
from mutual interference, users have incentive to choose their carriers orthogonally (exactly like in OFDMA systems).
\ei

\section{Implementation Issues}\label{sec:implem}

Although Prop. \ref{prop:secondary user-power} and Prop. \ref{prop:existence-se} guarantee SE existence, it is still not clear whether users will be able to calculate this equilibrium in a decentralized environment where only
partial/local information is available at the mobile terminal.
Consequently, our goal in this section will be to study implementation issues related to the converge to the equilibrium and its speed along with the sensing problem.
So far, we have assumed that the channels are static. If the channels fluctuate stochastically over time, the associated game still admits an equilibrium, but the learning process
is no more deterministic; just the same, by employing the theory of
stochastic approximation, it can be shown that users still converge to
equilibrium \cite{panayotis_learning_jsac12}. In the next section, we propose a temporal difference learning algorithm that ensures convergence to the SE within a limited time.

\subsection{Learning-based approach}\label{sec:learn}

The interaction between the PU and the SU provides a
potential incentive for both agents to make decision process based
on their respective perceived payoff. Determining the equilibrium strategy of both the primary and the secondary
users requires in practice the knowledge of several informations which can not
be observed in a realistic scenario \cite{Fudenberg1998}. We propose, in this section, an
on-policy learning-based algorithm that allow the PU and the
SU to determine their strategies on-the-fly. Machine learning is a powerful technique where
learning is accomplished by real-time interactions with the environment, and proper utilization
of past experience.
In particular, we consider a well-known temporal difference learning where each user maintains state-value functions as a lookup tables in order to determine the optimal action in the current time slot \cite{Sutton98}. To cope with the hierarchical decision process between the PU and the SU, we further set an iteration scale parameter $N_{iter}$ which traduces how frequent the SU updates its state-value function and set new values of powers with respect to the PU. The PU's state-value function $q(\textbf{g},\textbf{p})$ is given by
\begin{eqnarray*}\label{Learning.pu}
q(\textbf{g}^{t-1},\textbf{p}^{t-1})\leftarrow (1-\beta_t)q(\textbf{g}^{t-1},\textbf{p}^{t-1})+\
\beta_t(u_1+\kappa q(\textbf{g}^{t},\textbf{p}^{t})),
\end{eqnarray*}
whereas, the SU's state-value function $Q(\textbf{g},\textbf{p})$ is
\begin{eqnarray*}\label{Learning.su}
Q(\textbf{g}^{t-1},\textbf{p}^{t-1})\leftarrow (1-\alpha_t)Q(\textbf{g}^{t-1},\textbf{p}^{t-1})+\alpha_t(u_2+\kappa Q(\textbf{g}^{t},\textbf{p}^{t})),
\end{eqnarray*}
where $\kappa$ is the discount factor, and $\beta_t$ and $\alpha_t$ are the learning rate factors satisfying
$\sum_{t=1}^\infty \beta_t=\infty$ and $\sum_{t=1}^\infty(\beta_t)^2<\infty$, respectively $\sum_{t=1}^\infty\alpha_t=\infty$ and $\sum_{t=1}^\infty(\alpha_t)^2<\infty$.

The pseudo-code for the proposed algorithm is given in Algorithm 1. Specifically, we consider an effective balancing between exploration and exploitation. Note that with a probability $\epsilon$ we explore new actions, while we choose the already established action with a probability $1-\epsilon$. Indeed, the trade-off between exploration and exploitation remains a challenging issue in stochastic learning process.\\

\textbf{Algorithm 1: Learning-based Algorithm for Energy Efficient Cognitive Radio Networks.}
\begin{algorithm}
 \label{lear_algo}
Initialize $q(\textbf{g},\textbf{p})=0$ and $Q(\textbf{g},\textbf{p})=0$ for all channel gains and transmit powers\;
Initialize $R$, $\textbf{g}_1$, $\textbf{g}_2$,$\textbf{p}_1$ and $\textbf{p}_2$;\\
 \While{true}{
 $\textbf{g}_1^{prev}=\textbf{g}_1$\;
 $\textbf{p}_1^{prev}=\textbf{p}_1$\;
 Observe the new channel gains $\textbf{g}_1=(g_{11},g_{12})$\;
 Select transmit power vector $\textbf{p}_1=(p_{11},p_{12})$ as follows\:
 $\textbf{p}_1=\arg\max\limits_{\textbf{p'}}q(\textbf{g}_1,\textbf{p'})$ with probability $(1-\epsilon)$, else choose a random transmit power vector\;
 \For{$n=1\rightarrow N_{iter}$}{
$\textbf{g}_2^{prev}=\textbf{g}_2$\;
$\textbf{p}_2^{prev}=\textbf{p}_2$\;
Observe the new channel gains $\textbf{g}_2=(g_{21},g_{22})$\;
Select transmit power vector $\textbf{p}_2=(p_{21},p_{22})$ as follows\:
$\textbf{p}_2=\arg\max\limits_{\textbf{p'}}Q(\textbf{g}_2,\textbf{p'})$ with probability $(1-\epsilon)$, else choose a random power vector\;
Use the transmit power vector $\textbf{p}=(p_1,p_2)$ and observe the reward $u_2$, and $u_1$ given by Eq. (\ref{eq:util-mc})\; $Q(\textbf{g}^{prev},\textbf{p}^{prev})\leftarrow (1-\alpha_t)Q(\textbf{g}^{prev},\textbf{p}^{prev}) +\alpha_t(u_2+\kappa Q(\textbf{g}_2,\textbf{p}_2))$\;
 $R=R+u_1$\;}
$q(\textbf{g}^{prev},\textbf{p}^{prev})\leftarrow (1-\beta_t)q(\textbf{g}^{prev},\textbf{p}^{prev}) +\beta_t(R+\kappa q(\textbf{g}_1,\textbf{p}_1))$\; $R=0$\;
 }
\end{algorithm}

The following proposition proves that the learning-based algorithm for energy efficient cognitive radio networks converges to the optimal policy.
\begin{prop}\label{prop:convergence}
The learning-based algorithm converges w.p.1 to the optimal $Q$-function.
\end{prop}
The proof of Prop. \ref{prop:convergence} is given in Appendix \ref{app:convergence}. The learning rate time is addressed in the following proposition.
\begin{prop}\label{prop:convspeed}
Let $Q_T$ and $q_T$ be the value of the learning-based algorithm for the SU and the PU respectively. Then, we have $max\{||Q_T-Q^*||,||q_T-q^*||\}\leq\epsilon$ with probability at least $1-\delta$, given that

{\small{\beq
T=\ds\Omega \left( N_{iter}\cdot(L+\Phi\cdot L+1)^{\frac{1}{\beta}\cdot \ln(\frac{V_{max}}{\epsilon})}\cdot\frac{V_{max}^2 \ln(\frac{|\mathbb{S}||\mathbb{A}|V_{max}}
{\delta\beta\epsilon\Phi})}{(\Phi\beta\epsilon)^2}\right)
\eeq}}
where $\beta=(1-\kappa)/2$, $V_{max}=\frac{R_{max}}{1-\kappa}$, $R_{max}$ is the maximum reward obtained, and $|\mathbb{S}|$ and $|\mathbb{A}|$ are the number of possible states and strategies respectively. For a sequence of state-action pairs let the covering time, denoted by $L$, be an upper bound on the number of state-action pairs starting from any pair, until all state-action appear in the sequence. Indeed, the convergence speed of the proposed algorithm depends on the iteration scale parameter $N_{iter}$. The notation $T=\Omega(\Psi)$ implies that there are constants $c_1$ and $c_2$ such that $T\geq c_1 f \ln^{c_2}(\Psi)$.
\end{prop}
The proof of Prop. \ref{prop:convspeed} is given in Appendix \ref{app:convspeed}.

\subsection{Spectrum Sensing }\label{sec:sens}

In the current Stackelberg model, Proposition \ref{prop:coordination} claims that
the SU transmits over a certain frequency carrier in order to reach $\gamma^*$ only when the PU does not.
This enables public access to the new spectral ranges without sacrificing the
transmission quality of the actual license owners. Typically, the PU comes first in the system, estimates his channel gains $g_{1k}$ over his two carriers and adapts his transmit power using Prop. \ref{prop:existence-se}.
The SU comes later in the system, estimates his channel
links $g_{2k}$ over his two carriers and chooses his transmit power using Prop. \ref{prop:secondary user-power}. Such an assumption could be further
justified by the fact that in an asynchronous context, the probability that
two users decide to transmit at the same moment is negligible as the number
of users is limited. Thus, within this setting, the PU is assumed
to be oblivious to the presence of the SU. The PU communicates
with his BS while the SU listens to the wireless channel. The SU has only to
reliably detect the carrier used by the PU and not the PU's transmit power as
it is the case in the single carrier context in \cite{samson-twc09}).
Many well-known techniques
were developed in order to detect the holes in the spectrum band (energy
detection~\cite{Urkowitz67}, feature detection~\cite{Giannakis94}, etc.).

\section{Numerical illustration}\label{sec:sim}

In this section, we present a comprehensive Matlab-based simulation of the
CRN described in the previous sections. We consider the energy efficiency
function proposed in most papers dealing with power control games that is
$f(x)=(1-e^{-x})^M$, where $M=100$ is the block length in bits. This results on $\gamma^* = 6.4$ ($= 8.1$ dB).
$\mbox{SNR}=1/\sigma^2$ and the rate $R_n =1$ Mbps for $n=\{1,2\}$.

\begin{figure}[t]
\centering
\vspace*{0.5cm}
\hspace*{-0cm}
\centering
\includegraphics[height =4cm,width=8.5cm]{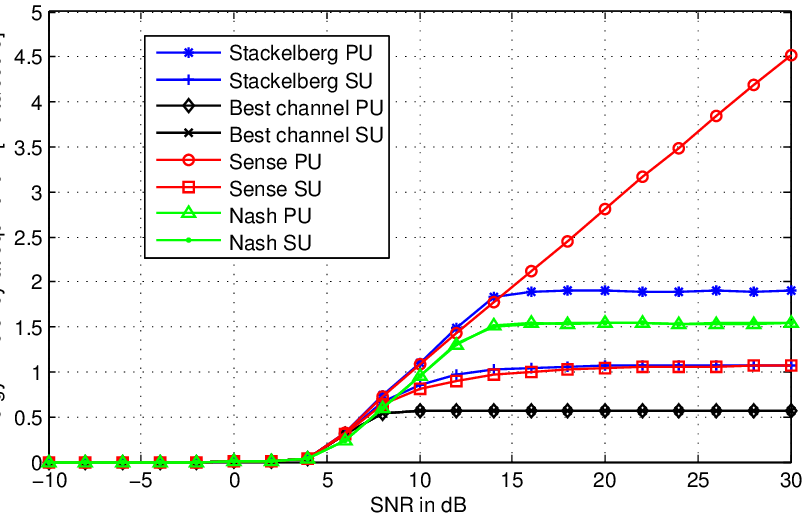}
\vspace*{-0cm}
\caption{Energy efficiency at the equilibrium as function of the $\mbox{SNR}$ for different schemes.}
\label{fig:ee_snr}
\vspace*{0.5cm}
\hspace*{-0cm}
\centering
\includegraphics[height =4cm,width=8.5cm]{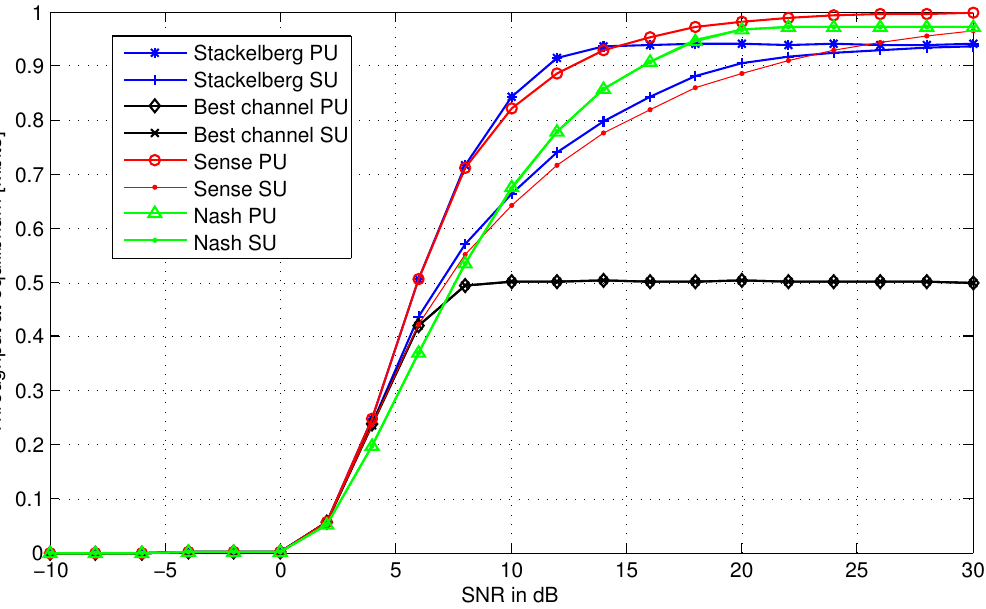}
\vspace*{-0cm}
\caption{Throughput at the equilibrium as function of the $\mbox{SNR}$ for different schemes.}
\label{fig:th_snr}
\end{figure}

\subsection{Energy Efficiency as a function of the SNR}

This section is devoted to performance comparison of the proposed Stackelberg
scheme with respect to traditional schemes. As far as sum energy efficiency
comparison is concerned, this can be conducted by considering the four
following schemes:

\bi
\item \emph{the Stackelberg model}: the one proposed in this paper,
\item \emph{the Nash model}: each user chooses his power level according to
 \cite{meshkati-jsac-2006},
\item \emph{the best channel model}: each user chooses to transmit on his "best"
 channel (i.e., the one with the best channel gain) without sensing,
\item \emph{the best channel with sensing}: the PU chooses the "best"
 channel to transmit on. The SU senses the spectrum and
 transmits on the vacant sub-band. Here we assume perfect sensing of the
 idle sub-band by the SU.
 \ei

In Figure \ref{fig:ee_snr}, we plot the energy efficiency at equilibrium as
function of the \mbox{SNR}. Interestingly, we see that the energy efficiency
of the PU at the SE performs the same than in
the sensing scenario till $12$ dB, while the energy efficiency of the
SU at the SE is always the same than in the
scheme where sensing is done by the SU.
Moreover, the Stackelberg model outperforms all the other strategies. This is
due to the Stackelberg mechanism in which the PU anticipates the
SU's action. In particular, we found out that the PU
achieves an energy efficiency gain up to $80\%$ with respect to the Nash
strategy at $12$ dB. As expected, results in Figure \ref{fig:ee_snr} also show
that the energy efficiency for the SU at SE is less than the one obtained at NE. This is due to
the fact that in Nash model, the PU does not anticipate the SU's
action. Notice that, as the $\mbox{SNR}$ decreases, all configurations tend
towards having the same (zero) energy efficiency. This can be justified by
the fact that, at low $\mbox{SNR}$ regime, whatever the power control
strategy each user chooses, the signal is overwhelmed by the noise.

Figure \ref{fig:th_snr} depicts the throughput at the equilibrium. We observe
approximately the same observations than in Figure \ref{fig:ee_snr}. Of
particular interest is the fact that the PU still outperforms all the other
strategies till $\mbox{SNR}=15$ dB whereas the throughput of the SU at
the SE is still less than the one obtained at the Nash
equilibrium. That is, the proposed Stackelberg scheme achieves a flexible
and desirable trade-off between energy efficiency and throughput maximization.

\subsection{Learning the Equilibria}

To proceed further with the analysis, we resort to simulate how the PU and
the SU users converge to the equilibria according to Algorithm
\ref{lear_algo} presented in Section \ref{sec:learn}. The noise variance is $\sigma^2=0.1$ which corresponds to a $\mbox{SNR} = 10$ dB. We consider an iteration scale $N_{iter}=10$, which means that the SU runs $10$ iterations for $1$ iteration of the PU.

\subsubsection{Static Channels}

\begin{figure}[t]
\vspace*{0.2cm}
\hspace*{-0.5cm}
\centering
\includegraphics[height =4.5cm,width=10cm]{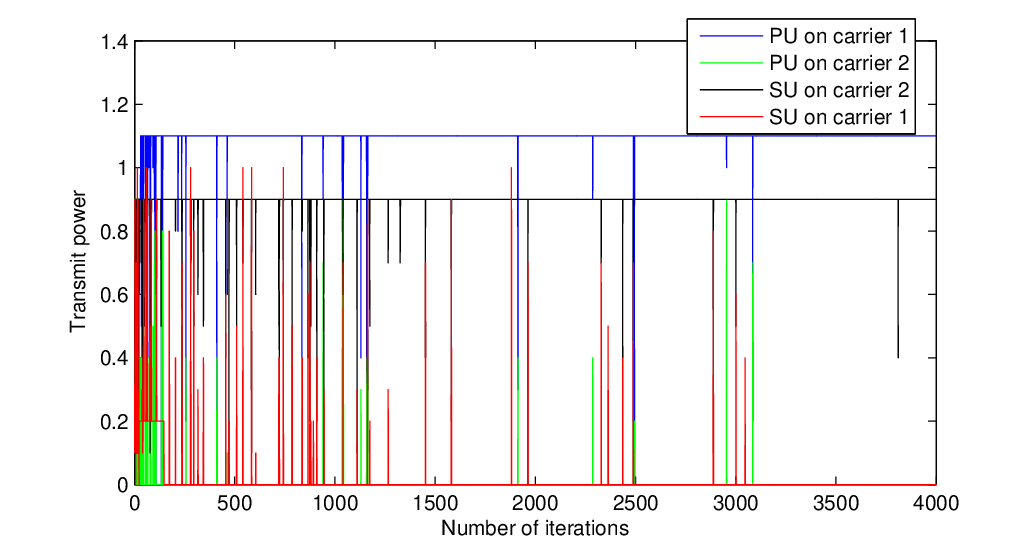}
\vspace*{-0.7cm}
 \caption{The transmit power at the Stackelberg equilibrium for both PU and SU when $\textbf{g}_1=(0.4,0.3)$ and $\textbf{g}_2=(0.6,0.5)$.} \label{fig:learnp1}
\vspace*{0.2cm}
\hspace*{-0.5cm}
\centering
\includegraphics[height =4.5cm,width=10cm]{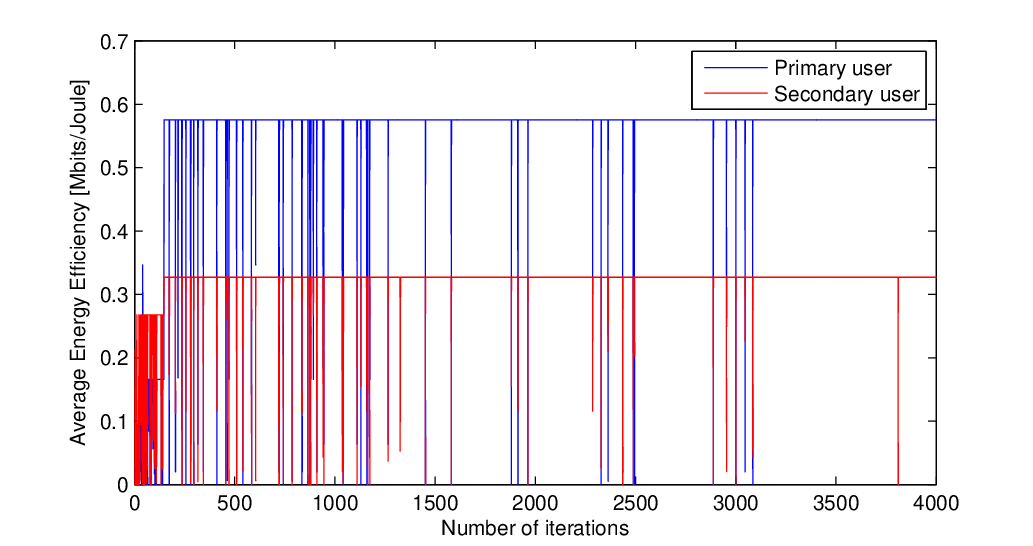}
\vspace*{-0.7cm}
 \caption{The energy efficiency at the Stackelberg equilibrium for both PU and SU when $\textbf{g}_1=(0.4,0.3)$ and $\textbf{g}_2=(0.6,0.5)$.} \label{fig:learnu1}
\end{figure}

In Figures \ref{fig:learnp1} and \ref{fig:learnu1}, we consider
static channel gains $g_{11}=0.4$, $g_{12}=0.3$, $g_{21}=0.6$ and $g_{22}=0.5$. We observe from Figure \ref{fig:learnp1} that the optimal power control decision of the PU is to transmit on the first
carrier whereas the SU chooses to transmit on the second carrier as claimed by Prop. \ref{prop:coordination}. Indeed, we have $\frac{g_{11}}{g_{12}}\simeq1.3\geq1$ and $\frac{g_{21}}{g_{22}}=1.2$ which is in the interval
$[\frac{1}{1+\gamma^*},1+\gamma^*]$. This means that the SE is given by Prop. \ref{prop:coordination}-a-ii yielding the following SE: $\left(\widetilde{p}_{11},\widetilde{p}_{12},\widetilde{p}_{21},\widetilde{p}_{22}\right)\simeq\left(1.1,0,0,0.9\right).
$

\begin{figure}[t]
\vspace*{-0cm}
\hspace*{-0.9cm}
\centering
\includegraphics[height =4.5cm,width=10.5cm]{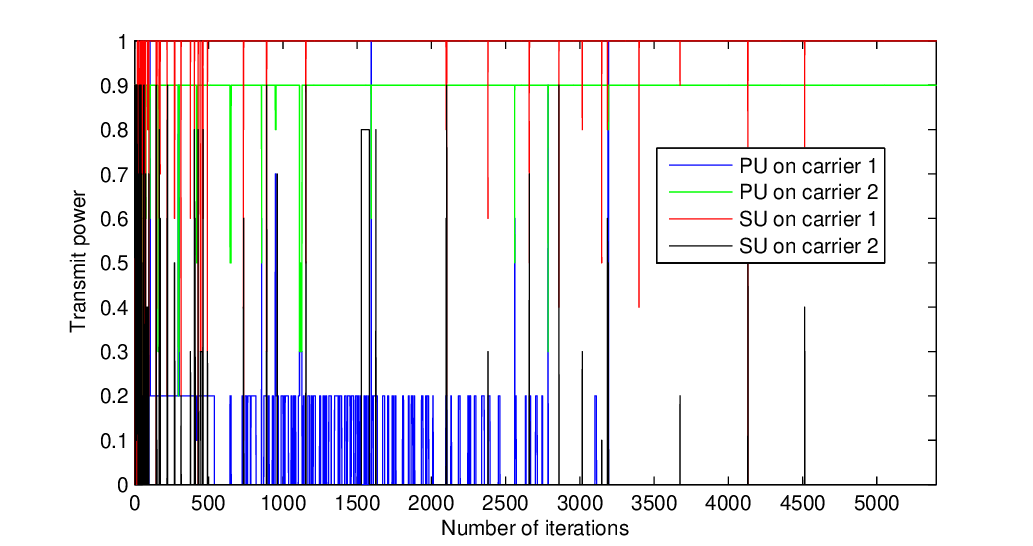}
\vspace*{-0.5cm}
\caption{The transmit power at the Stackelberg equilibrium for both PU and SU when $\textbf{g}_1=(0.2,0.3)$ and $\textbf{g}_2=(4,0.5)$.} \label{fig:learnp2}

\vspace*{0.2cm}
\hspace*{-0.5cm}
\centering
\includegraphics[height =4.5cm,width=10cm]{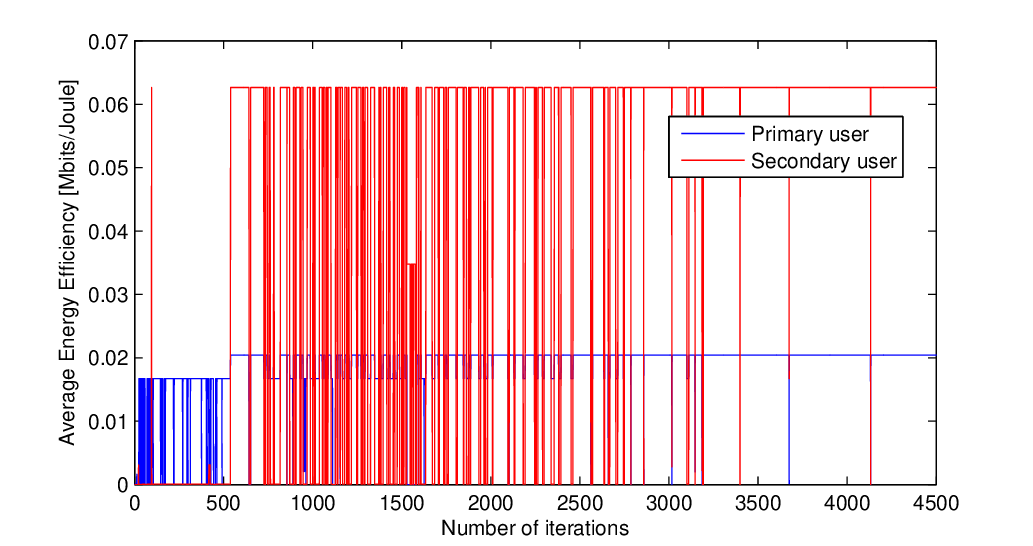}
\vspace*{-0.5cm}
\caption{The energy efficiency at the Stackelberg equilibrium for both PU and SU when $\textbf{g}_1=(0.2,0.3)$ and $\textbf{g}_2=(4,0.5)$.} \label{fig:learnu2}
\end{figure}

In Figure \ref{fig:learnp2} and \ref{fig:learnu2}, we change the second carrier's PU channel gain
to $g_{11}=0.2$ and the second carrier's SU channel gain to $g_{21}=4$. The SE changes accordingly. In fact, we have that $\frac{g_{11}}{g_{12}}\simeq 0.6<1$ and $\frac{g_{21}}{g_{22}}>1+\gamma^*$ which corresponds to the case (b-i) of Prop. \ref{prop:coordination} where the PU decides to transmit on the second carrier and the SU transmit on the first carrier yielding the following SE: $\left(\widetilde{p}_{11},\widetilde{p}_{12},\widetilde{p}_{21},\widetilde{p}_{22}\right)\simeq\left(0,0.9,1,0\right).
$

In Figure \ref{fig:learnu1} and \ref{fig:learnu2}, we look at the energy efficiency of the PU and the SU. In general case, the PU outperforms the SU since the PU anticipates the SU's action (see Fig. \ref{fig:learnu1}). However, it is illustrated in Fig. \ref{fig:learnu2} that, although he plays first, the PU performs worse that the SU at the equilibrium as the best SU's carrier ($g_{21}=4$) is much better than the PU's best carrier ($g_{12}=0.3$).

\subsubsection{Fading Channels}

\begin{figure}[t]
\vspace*{0.2cm}
\hspace*{-0cm}
\centering
\includegraphics[height =4.3cm,width=8.5cm]{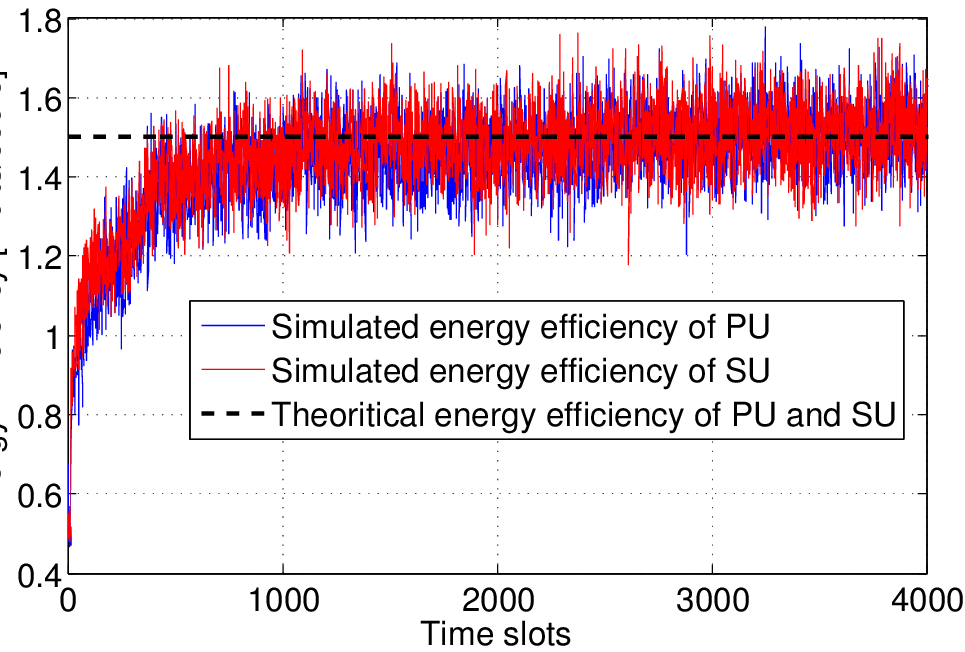}
\vspace*{-0.1cm}
\caption{The energy efficiency at the Nash equilibrium for both PU and SU.}
\label{fig:nash}

\vspace*{0.7cm}
\hspace*{-0cm}
\centering
\includegraphics[height =4.3cm,width=8.5cm]{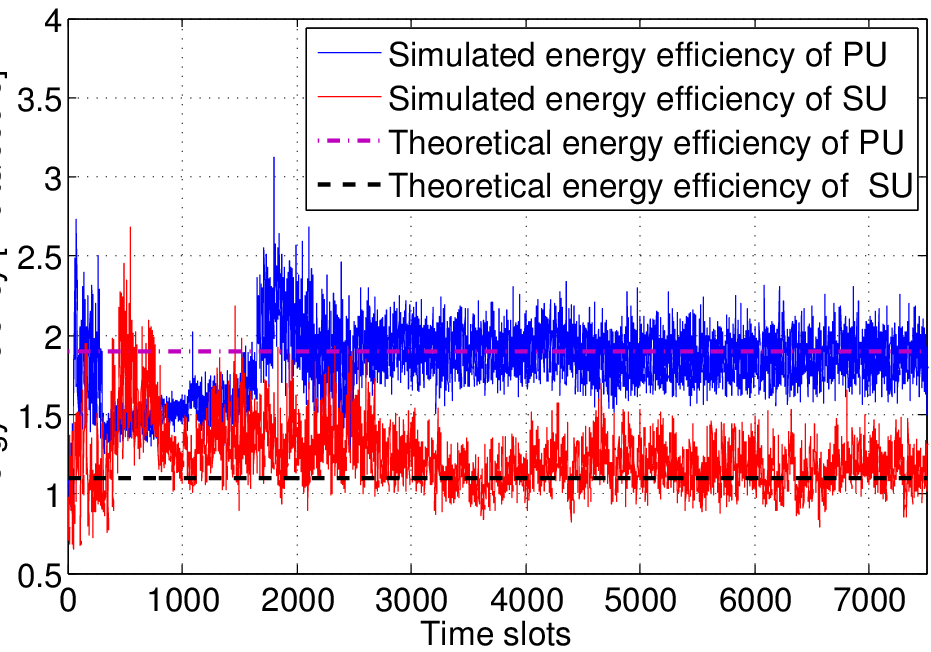}
\vspace*{-0.1cm}
\caption{The energy efficiency at the Stackelberg
equilibrium for both PU and SU.} \label{fig:stack}
\end{figure}

In Figure \ref{fig:nash}, we plot the energy efficiency of the PU and the SU
at the NE proposed in \cite{meshkati-jsac-2006} depending on
time. It is clear that both the PU and the SU converge to the same energy
efficiency since the Nash game is a one-shot game. We also observe that both
the PU and the SU converge to exactly the same energy efficiency of $1.6$
Mbit/Joule than the one obtained in Figure \ref{fig:ee_snr} at $\mbox{SNR} =10$ dB.
Next, we plot in Figure \ref{fig:stack}, the convergence of the energy
efficiency at the SE for both the PU and the SU. Again
we observe that the PU and the SU converge to the same energy efficiency of
$1.9$ Mbit/Joule and of $1.1$ Mbit/Joule respectively obtained in Figure
\ref{fig:ee_snr} at $\mbox{SNR} = 10$ dB. Moreover, as expected, that the
energy efficiency at the SE of the PU is higher than the
energy efficiency of the SU. Note that the variance of energy efficiency in Figures \ref{fig:nash} and \ref{fig:stack} is due to the fact that the fading channel states of the PU $\textbf{g}_1$ and the SU $\textbf{g}_2$ vary every time slot. Though, the algorithm still converges to
the equilibrium of an averaged game whose payoff
functions correspond to the users' achievable ergodic rates.\\\\

\section{Conclusion}
\label{sec:conc}

In this paper, we have proposed a hierarchical concept in a power control game
for energy efficient multi-carrier cognitive radio systems. We have firstly completely and analytically characterized the
Stackelberg equilibrium of such a game. Interestingly, we have shown that, although we have considered that each user is prone to interference from
the other transmitter on the same carrier, for the vast majority of cases, there exists a natural coordination pattern where the PU and the SU have incentive to choose their transmitting carriers orthogonally (like in OFDMA systems). The proposed system
goes toward the vision of a fully coordinated cognitive radio multi-carrier network,
whereby transmit powers are coordinated across the users. Then, we have compared the users' energy efficiency of the proposed hierarchal game with those obtained in a
standard non-cooperative setting. In addition to allowing coordination of the
spectrum usage, the proposed power control game provides
additional functionalities that can be used in energy efficient CRN. In particular, the proposed Stackelberg scheme achieves a flexible
and desirable trade-off between energy efficiency and throughput maximization.
For implementation purposes, the SU has only to reliably sense the spectral environment (and not the PU's transmit power as it is the case in the single carrier context in \cite{samson-twc09}) and then decides to transmit only on the \emph{best} carrier left idle by the PU.
Finally, with extensive measurement-driven
simulations we show that the proposed game model converges to the desired
equilibria in a small number of steps, and hence are amenable to
practical implementation.

\bibliographystyle{IEEEtran}
\bibliography{C:/Users/mhaddad/Dropbox/mybib}

\appendix

\subsection{Proof of Prop. \ref{prop:existence-se}: \textbf{Existence and uniqueness of the PU's power control at the SE}}\label{app:exis}

\begin{proof}
Given Proposition \ref{prop:secondary user-power}, we have that the power control
vector of the SU in Region ${\mathcal{A}}$ and ${\mathcal{B}}$ are given, respectively, by
$$
\mathbf{p_2^{\mathcal{A}}}(p_{12})=(0,\frac{\gamma^{*}(\sigma^2+g_{12} p_{12})}{g_{22}})
$$
 and
$$
\mathbf{p_2^{\mathcal{B}}}(p_{11})=(\frac{\gamma^{*}(\sigma^2+g_{11} p_{11})}{g_{21}},0).
$$
Based on the above equations, we can compute the explicit expression of the
PU's SINR on each carrier for both regions, namely
\begin{equation*}
\gamma_{11}= \left\{\begin{array}{lr}
\ds\frac{g_{11}p_{11}}{\sigma^2},&\quad \mbox{in Region ${\mathcal{A}}$},\\
\ds\frac{g_{11}p_{11}}{\sigma^2(1+\gamma^*)+\gamma^*g_{11}p_{11}},& \quad \mbox{in Region ${\mathcal{B}}$}
\end{array}
\right.
\end{equation*}
\begin{equation*}
\gamma_{12}= \left\{\begin{array}{lr}
\ds\frac{g_{12}p_{12}}{\sigma^2(1+\gamma^*)+\gamma^*g_{12}p_{12}},&\quad \mbox{in Region ${\mathcal{A}}$},\\
\ds\frac{g_{12}p_{12}}{\sigma^2},& \quad \mbox{in Region ${\mathcal{B}}$}
\end{array}
\right.
\end{equation*}
It follows that the utility function of the PU given by Equation
(\ref{eq:util-mc}) for Region ${\mathcal{A}}$ can be expressed as
\begin{eqnarray*}
u_1^{\mathcal{A}}(p_{11},p_{12})&=&\frac{R_1f(\gamma_{11})+R_1f(\gamma_{12})}{p_{11}+p_{12}},\\
&=&\frac{R_1f(\frac{g_{11}p_{11}}{\sigma^2})+R_1f(\frac{g_{12}p_{12}}{\sigma^2(1+\gamma^*)+\gamma^*g_{12}p_{12}})}{p_{11}+p_{12}}
\end{eqnarray*}
Similarly, in Region ${\mathcal{B}}$, the PU's utility function is
$$
u_1^{\mathcal{B}}(p_{11},p_{12})=\frac{R_1f(\frac{g_{11}p_{11}}{\sigma^2(1+\gamma^*)+\gamma^*g_{11}p_{11}})+R_1f(\frac{g_{12}p_{12}}{\sigma^2})}{p_{11}+p_{12}}
$$

Without loss of generality, the analysis is given only for Region ${\mathcal{A}}$.
Similar approach can be adopted for Region ${\mathcal{B}}$. We first derive the utility
of the PU $u_1^{\mathcal{A}}(p_{11},p_{12})$ w.r.t $p_{11}$. We obtain

{{ \small{\beqq\label{eq:der1} \frac{\partial u_1^{\mathcal{A}}(p_{11},p_{12})}{\partial p_{11}}
= R_1\cdot\frac{ f^{\prime}(\gamma_{11})\cdot \frac{g_{11} }{\sigma^2} \cdot
(p_{11}+p_{12})-[f(\gamma_{11})+ f(\gamma_{12})]}{(p_{11}+p_{12})^2}
\eeqq}}}

Now, let us compute the derivative of the PU's utility $u_1^{\mathcal{A}}$ on
the Region ${\mathcal{A}}$ w.r.t $p_{12}$. We have

{\small{\beqq\label{eq:der2} \frac{\partial u_1^{\mathcal{A}}(p_{11},p_{12})}{\partial p_{12}}
= R_1\cdot\frac{ f^{\prime}(\gamma_{12})\cdot \frac{\partial
\gamma_{12}}{\partial p_{12}} \cdot (p_{11}+p_{12})-[ f(\gamma_{11})+
f(\gamma_{12})]}{(p_{11}+p_{12})^2} \eeqq}}

where $\ds\frac{\partial
\gamma_{12}}{\partial p_{12}}=\frac{g_{12} (\sigma^2+\gamma^*
(\sigma^2+p_{12} g_{12}))-g_{12} p_{12} \gamma^*
g_{12}}{\left[\sigma^2+\gamma^* (\sigma^2+p_{12} g_{12})\right]^2}$. Knowing that $\ds\sigma^2+p_{12} g_{22}=\frac{\sigma^2
(1+\gamma_{12})}{1-\gamma^* \gamma_{12}}$ and after some simple
simplifications, we obtain that $\ds\frac{\partial \gamma_{12}}{\partial
p_{12}}= \frac{g_{12}(1-\gamma_{12} \gamma^*)^2}{\sigma^2 (1+\gamma^*)}$.

We shall now look for a couple $\ds(p_{11},p_{12})$ such that $\ds\frac{\partial
u_1^{\mathcal{A}}}{\partial p_{11}}(p_{11},p_{12})=\frac{\partial u_1^{\mathcal{A}}}{\partial
p_{12}}(p_{11},p_{12})=0$. It follows from the above results that a couple
$(p_{11},p_{12})$ is solution of the following system

{\small{\begin{equation*}
(\mathcal{S}):\left\{\begin{array}{lr}
\displaystyle f^{\prime}(\gamma_{11})\frac{g_{11}}{\sigma^2}(p_{11}+p_{12})&=f(\gamma_{11})+f(\gamma_{12})\\
\displaystyle f^{\prime}(\gamma_{12})\frac{g_{12}(1-\gamma_{12}\gamma^*)^2}{\sigma^2(1+\gamma^*)}(p_{11}+p_{12})&=f(\gamma_{11})+f(\gamma_{12})
\end{array}
\right.
\end{equation*}}}
with $\ds\gamma_{11}=\frac{g_{11}p_{11}}{\sigma^2}$ and
$\ds\gamma_{12}=\frac{g_{12}p_{12}}{\sigma^2(1+\gamma^*)+\gamma^*g_{12}p_{12}}$.

The solutions of the above system are given by \beq\label{eq:p11-stack-A}
p_{11}=\frac{\sigma^2 \gamma_{11}}{g_{11}} \eeq and \beq\label{eq:p12-stack-A}
p_{12} = \frac{\sigma^2}{g_{12}}\frac{{\gamma_{12}}
(1+\gamma^*)}{(1-\gamma^* {\gamma_{12}})}.
\eeq

In Region ${\mathcal{A}}$, Eq.~(\ref{eq:A}) yields to the following relation between the powers of the
PU:
\begin{eqnarray}\label{rA}
p_{12} \leq p_{11}\frac{g_{11} g_{22}}{g_{12} g_{21}} + \sigma^2
\frac{(g_{22}-g_{21})}{g_{12} g_{21}}
\end{eqnarray}
which means that for all $p_{11}> 0$, the PU's power on the second carrier
$p_{12}$ is in the interval $\left[0,p_{11}\frac{g_{11} g_{22}}{g_{12} g_{21}}
+ \sigma^2 \frac{(g_{22}-g_{21})}{g_{12} g_{21}}\right]$. Therefore, our
problem boils down to show that, for a fixed $p_{11}$, the partial derivative
of $u_1^{\mathcal{A}}(p_{11},p_{12})$ w.r.t. $p_{12}$ in the neighboring of zero is a
strictly decreasing function. The limit of the partial derivative of
$u_1^{\mathcal{A}}(p_{11},p_{12})$ when $p_{12}$ tends to zero is given by
$$
\forall p_{11}> 0, \quad \lim_{p_{12} \rightarrow 0^+}\frac{\partial
u_1^{\mathcal{A}}(p_{11},p_{12})}{\partial
p_{12}}=\frac{-R_1f(\frac{g_{11}p_{11}}{\sigma^2})}{(p_{11})^2}<0,
$$
where we used from \cite{rodriguez-globecom-2003} the fact that
$f(0)=f^{\prime}(0)=0$ yielding that $\widetilde{p}_{12}=0$ in Region $\mathcal{A}$. So far, we have proved that maximizing the utility of the PU in Region $\mathcal{A}$ implies
maximizing this utility function by considering that $p_{12}=0$. Then, Condition (\ref{rA}) becomes
$$
p_{11}\geq \frac{\sigma^2(g_{21}-g_{22})}{g_{11}g_{22}}:=\widehat{p}_{11}.
$$
On the other hand, we know that the function $u_1(p_{11},0)$ is maximized for
$\widetilde{p}_{11}=\frac{\sigma^2\gamma^*}{g_{11}}$. It follows
that, if $\frac{\sigma^2\gamma^*}{g_{11}}\geq\widehat{p}_{11}$ (i.e.,
$\frac{g_{22}}{g_{21}}\geq\frac{1}{1+\gamma^*}$), then the utility of the PU in
Region $\mathcal{A}$ is maximized when $\widetilde{p}_{11}=\frac{\sigma^2\gamma^*}{g_{11}}$ and $\widetilde{p}_{12}=0$. Otherwise, if
$\frac{\sigma^2\gamma^*}{g_{11}}<\widehat{p}_{11}$, the utility of the PU in Region ${\mathcal{A}}$ is maximized when $\widetilde{p}_{11}=\widehat{p}_{11}$ and $\widetilde{p}_{12}=0$.

In Region ${\mathcal{B}}$, the same methodology is adopted by replacing $p_{11}=0$ in Eq.~(\ref{eq:B}). We end up with the condition below
$$
p_{12}\geq \frac{\sigma^2(g_{22}-g_{21})}{g_{12}g_{21}}:=\widehat{p}_{12}.
$$
Moreover, we know that the function $u_1(0,p_{12})$ is maximized for
$\widetilde{p}_{12}=\frac{\sigma^2\gamma^*}{g_{12}}$. It follows
that, if $\frac{\sigma^2\gamma^*}{g_{12}}\geq\widehat{p}_{12}$ (i.e.,
$\frac{g_{22}}{g_{21}}\leq1+\gamma^*$), then the utility of the PU over Region ${\mathcal{B}}$ is
maximized when $\widetilde{p}_{12}=\frac{\sigma^2\gamma^*}{g_{12}}$ and $\widetilde{p}_{11}=0$. Otherwise, if
$\frac{\sigma^2\gamma^*}{g_{12}}>\widehat{p}_{12}$, the utility of the PU over Region ${\mathcal{B}}$ is
maximized when $\widetilde{p}_{12}=\widehat{p}_{12}$ and $\widetilde{p}_{11}=0$.
\end{proof}

\subsection{Proof of Proposition \ref{prop:stack-0}}\label{app:stack-0}

\begin{proof}
Assume that the PU transmits over one carrier, say carrier $i$.
\begin{itemize}
\item If the SU does not transmit on the carrier $i$, i.e.,
 $L_2(\mathbf{p_1})=j \neq i$. Then, we have that
 $\ds p_{1i}=\frac{\sigma^2\gamma^*}{g_{1i}}$ yielding that
 $\ds \widehat{h}_{2j}=\frac{g_{2j}}{\sigma^2}>\widehat{h}_{2i}=\frac{g_{2i}}{\sigma^2(1+\gamma^*)}$
 which is equivalent to
$$
\frac{g_{2j}}{g_{2i}}>\frac{1}{1+\gamma^*},
$$
Then, we have proved that Condition
(\ref{chfol}) is sufficient.
\item If Condition (\ref{chfol}) is satisfied it means that there exists a
 carrier $j$ such that $g_{2j}>\frac{g_{2i}}{(1+\gamma^*)}$. Then, we
 assume that the SU transmits over $i$, which means that
$$
\widehat{h}_{2i}>\widehat{h}_{2j}=\frac{g_{2j}}{\sigma^2}.
$$
Suppose that the two players transmit over channel $i$ and the power used
by the PU at the SE is higher compared to the case
when a user is alone on a carrier \cite{samson-twc09}. Then, the power
$p_{1i}$ used by the PU is higher than $\frac{\sigma^2\gamma^*}{g_{1i}}$.
This implies that the effective carrier gain of the SU on the carrier $i$
is:
$$
\widehat{h}_{2i}=\frac{g_{2i}}{\sigma^2+g_{1i}p_{1i}}<\frac{g_{2i}}{\sigma^2(1+\gamma^*)}<\frac{g_{2j}}{\sigma^2}=\widehat{h}_{2j}.
$$
But this is in contradiction with the assumption that the SU transmits over
carrier $i$, then the SU does not transmit over carrier $i$ (the one chosen
by the PU). We have then proved the sufficient condition.
\end{itemize}
\end{proof}

\subsection{Proof of Prop. \ref{stackNash}}\label{app:stackNash}
\begin{proof}
We prove this proposition considering only Region $\mathcal{A}$ as it is the same
idea for Region ${\mathcal{B}}$. It is preferable for the PU to transmit over the
same carrier than the SU, the second one in this area, if and only if the
utility at the SE when the PU and the SU transmit on the
second carrier is higher than the utility of the PU when he is alone to
transmit over the first carrier.

The maximum utility for the PU, in Region $\mathcal{A}$, when he is alone to transmit
over the first carrier is given by:
$$
u_1^{NE}=\frac{R_1f(\gamma^*)}{\sigma^2\gamma^*}g_{11}.
$$

When both users transmit over the second carrier and the PU plays the Nash
action, i.e., $\ds p_{12}=\frac{\sigma^2\gamma^*}{g_{12}(1-\gamma^*)}$, the
best-response function of the SU is to choose the power
$\ds p_{22}=\frac{\sigma^2\gamma^*}{g_{22}(1-\gamma^*)}$. This NE
exists if the target SINR $\gamma^*$ is less than $1$. Then, the PU's utility at the NE is
$$
u_1^{NE}=\frac{R_1f(\gamma^*)(1-\gamma^*)}{\sigma^2\gamma^*}g_{12}.
$$
This result is true if the Nash action of the PU
$(p_{11},p_{12})=(0,\frac{\sigma^2\gamma^*}{g_{12}(1-\gamma^*)})$ is inside the
Region $\mathcal{A}$. This is true if and only if
$$
\frac{\sigma^2\gamma^*}{g_{12}(1-\gamma^*)}\leq\frac{\sigma^2(g_{22}-g_{21})}{g_{12}g_{21}},
$$
which is equivalent to
$$
\frac{g_{21}}{g_{22}}\leq1-\gamma^*.
$$
Thus, the PU's utility at the NE is better than the utility if
he transmits on the second carrier if and only if:
$$
\frac{g_{11}}{g_{12}}\leq1-\gamma^*.
$$
Then, if $\frac{g_{11}}{g_{12}}\leq1-\gamma^*$ and
$\frac{g_{21}}{g_{22}}\leq 1-\gamma^*$, we have that $u_1^{NE}>u_1^{SE}$. But, the
utility of the PU at a SE is, by definition, better or
equal than its utility if he plays the Nash action (the best-response function of the SU if the PU plays the Nash is the Nash). Then, if
$\frac{g_{11}}{g_{12}}\leq1-\gamma^*$ and $\frac{g_{21}}{g_{22}}\leq1-\gamma^*$ the
utility of the PU at the SE when the two players
transmit over the second carrier, is better than the utility of the PU if he
transmits alone on the first carrier.

We have similar analysis with Region ${\mathcal{B}}$, in which the SU transmits over
the first carrier. Over this region, the two players transmit over the first
carrier if and only if the following conditions are satisfied:
$$
\frac{g_{11}}{g_{12}}\geq\frac{1}{1-\gamma^*},\quad \mbox{and}\quad \frac{g_{21}}{g_{22}}\geq\frac{1}{1-\gamma^*}.
$$

\end{proof}

\subsection{Proof of Prop. \ref{prop:convergence}}\label{app:convergence}
\begin{proof}
The proposed algorithm is a two-time scale version of the well known $Q$-learning algorithm. Since both the utilities of the PU and the SU depend on the states and actions of PU and SU, i.e., $\textbf{g}$ and $\textbf{p}$, the utility functions $u_1$ and $u_2$ in Eq. (\ref{eq:util-mc}) are not deterministic, and considered as random variables instead. In fact, given the state $\mb g_2$ and the action $\mb p_2$ of the SU, the observed reward of the SU depends also on the state $\mb g_1$ and the action $\mb p_1$ of the PU, which are unknown for the SU.
The $Q$-learning algorithm for the SU given by
\begin{eqnarray*}
Q(\textbf{g}_2^{k-1},\textbf{p}_2^{k-1})&\leftarrow&(1-\alpha_k) Q(\textbf{g}_2^{k-1},\textbf{p}_2^{k-1}) \\&&+\alpha_k(u_2(\textbf{g}_2^{k-1},\textbf{p}_2^{k-1})+\kappa Q(\textbf{g}_{2k}\textbf{p}_{2k})),
\end{eqnarray*}
converges to the optimal $Q^*(\bf g_2,\bf p_2)$ value. In fact, since we have
\begin{itemize}
\item the state and action spaces are finite,
\item $\sum_{k=1}^\infty\alpha_k=\infty,\sum_{k=1}^\infty(\alpha_k)^2<\infty$ uniformly w.p.1,
\item $Var\{u_2(\textbf{g}_2,\textbf{p}_2)\}$ is bounded.
\end{itemize}
We obtain from Theorem 2 of \cite{Jaakkola94convergenceof} that the $Q$-learning algorithm for the SU converges.
Similarly, the $Q$-learning algorithm for the PU is expressed as
\begin{eqnarray*}
q(\textbf{g}_1^{k-1},\textbf{p}_1^{k-1})&\leftarrow& (1-\beta_k)q(\textbf{g}_1^{k-1},\textbf{p}_1^{k-1}) \\&&+\beta_k(u_1+\kappa q(\textbf{g}_{1k},\textbf{p}_{1k})),
\end{eqnarray*}
converges to the optimal $q^*(\bf g_1,\bf p_1)$ value.
This concludes the proof.
\end{proof}

\subsection{Proof of Prop. \ref{prop:convspeed}}\label{app:convspeed}
\begin{proof}
Let $Q_t$ and $q_t$be the value of the asynchronous $Q$-learning algorithm using linear learning (results for the polynomial learning rate exists also). Then, we obtain from Theorem 5 \cite{Even_Q-learning_03} that with probability $1-\delta$, for any positive constant $\Phi$ we have $\max\{||Q_t-Q^*||,||q_t-q^*|\}\leq\epsilon$, given that

{\small{\beq
T=\ds\Omega \left( N\cdot(L+\Phi\cdot L+1)^{\frac{1}{\beta}\cdot \ln(\frac{V_{max}}{\epsilon})}\cdot\frac{V_{max}^2 \ln(\frac{|S||A|V_{max}}
{\delta\beta\epsilon\Phi})}{(\Phi\beta\epsilon)^2}\right)
\eeq}}
\end{proof}

\end{document}